\documentclass[trackchanges,twocolumn]{aastex701}

\begin{document}

\title{Upper Limit on the Non-Gravitational Acceleration and Lower Limits on the Nucleus Mass and Diameter of 3I/ATLAS}

\author[orcid=0000-0003-2586-2697,gname=Richard, sname=Cloete]{Richard Cloete}
\affiliation{Institute for Theory \& Computation, Harvard University, 60 Garden Street, Cambridge, MA 02138, USA}
\email[show]{richard.cloete@cfa.harvard.edu}  

\author[orcid=0000-0003-4330-287X,gname=Abraham, sname=Loeb]{Abraham Loeb} 
\affiliation{Institute for Theory \& Computation, Harvard University, 60 Garden Street, Cambridge, MA 02138, USA}
\email[show]{aloeb@cfa.harvard.edu} 

\author[orcid=0000-0003-2586-2697,gname=Peter, sname=Veres]{Peter Vere\v{s}}
\affiliation{Minor Planet Center, Harvard \& Smithsonian Center for Astrophysics, 60 Garden Street, Cambridge, MA 02138, USA}
\email[]{peter.veres@cfa.harvard.edu}

\begin{abstract}
We use astrometric data on 3I/ATLAS compiled by the Minor Planet Center from May 15 to September 23, 2025, and derive an upper limit on statistically-significant deviation from the best-fit gravity-based trajectory. The residuals imply that the non-gravitational acceleration is smaller than $\sim 3\times 10^{-10}~{\rm au~d^{-2}}$. Based on the total mass loss rate and outflow speed inferred from JWST data on August 6, 2025, we derive lower limits on the mass and diameter of 3I/ATLAS of $3.3\times 10^{16}~{\rm g}$ and $5~{\rm km}$, respectively. 
\end{abstract}


\section{Introduction}

In this note, we analyze astrometric observations of the interstellar object 3I/ATLAS spanning the period of May 15 to September 23, 2025 (UTC). We adopt optical astrometry from the Minor Planet Center (MPC) and compare it to the sky coordinates for the best-fit gravity-based orbit.

4,022 astrometric observations reported by 227 distinct observatory codes were retrieved from the MPC database\footnote{\url{https://data.minorplanetcenter.net/explorer/}}, excluding the TESS astrometry~\citep{TESS}.

\section{Residual Analysis}
\label{subsec:residuals}
The orbit of 3I/ATLAS was computed with the MPC’s orbfit package \citep{orbfit2013}, adopting the MPC astrometric weighting scheme \citetext{\citeauthor{SpotoWeightsInPrep}, in preparation}. The solution assumed a gravity-only dynamical model including perturbations from all planets, the Earth–Moon system treated separately, and the 17 most massive minor planets. Planetary positions were taken from the DE441 ephemeris \citep{Park2021DE441}. The fit provides residuals for each astrometric observation.

The computed osculating Keplerian orbital elements of 3I/ATLAS with 1$\sigma$ uncertainties for the epoch MJD = 60885.672886722 TDT were as follows: 
$q= 1.3563 (\pm 0.0001)~{\rm au}$, 
$e=6.1386 \pm 0.0006$, $i=175.1130 (\pm 0.0001)~{\rm deg}$, 
$\Omega=322.1559 (\pm 0.0012)~{\rm deg}$,
$\omega=128.0111 (\pm 0.0008)~{\rm deg}$,
$T_p({\rm MJD~TDT})=60977.483 (\pm 0.0004)$.

For the period of July 1 to September 23, 2025, the RA and Dec residuals set an upper limit on the mean shift of $0.025 (\pm 0.028)^{\prime\prime}$ and $0.019 (\pm 0.02)^{\prime\prime}$, respectively.  

\begin{figure*}[htbp]
    \centering
    \includegraphics[width=0.48\textwidth]{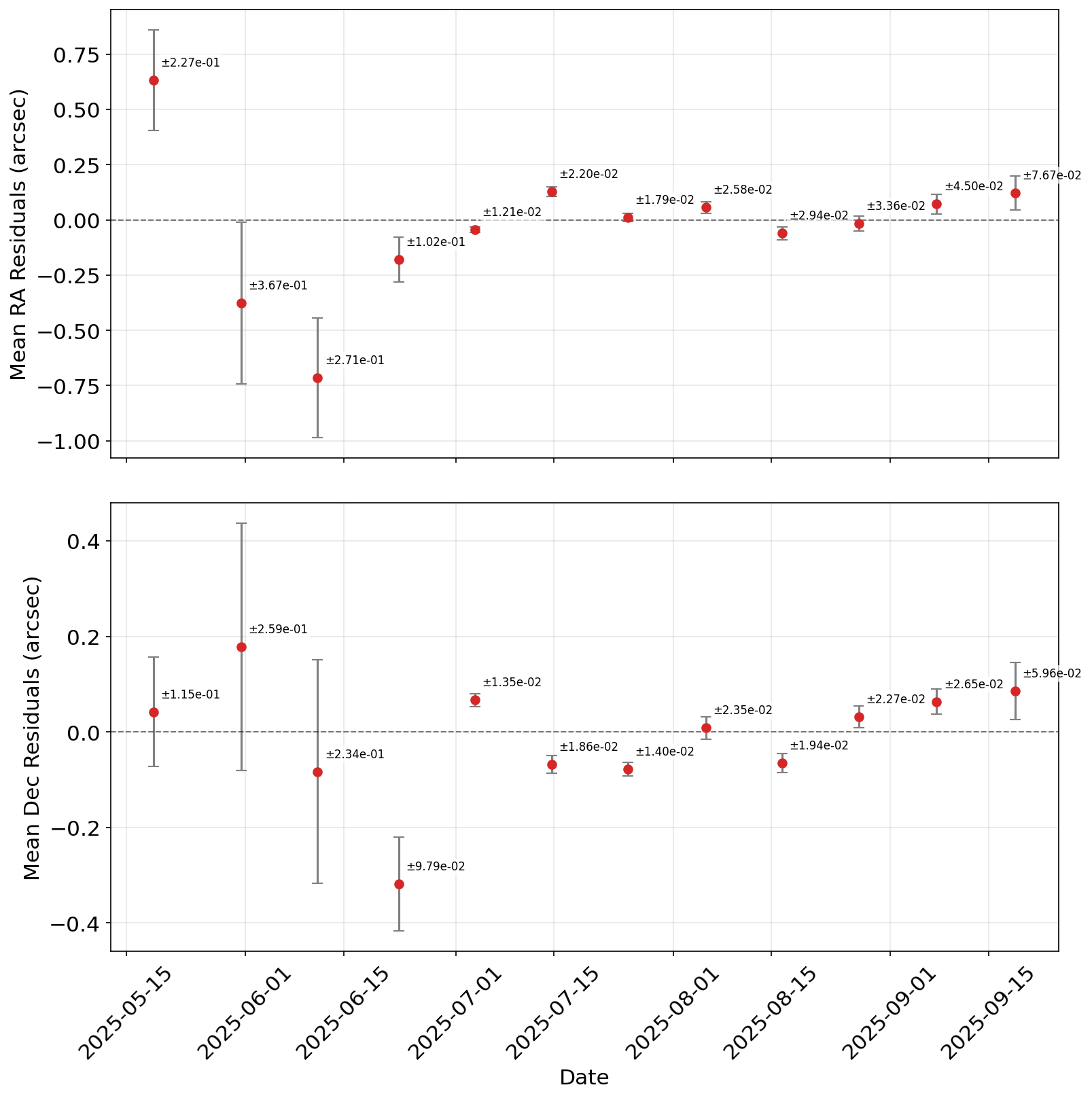}
    \hfill
    \includegraphics[width=0.48\textwidth]{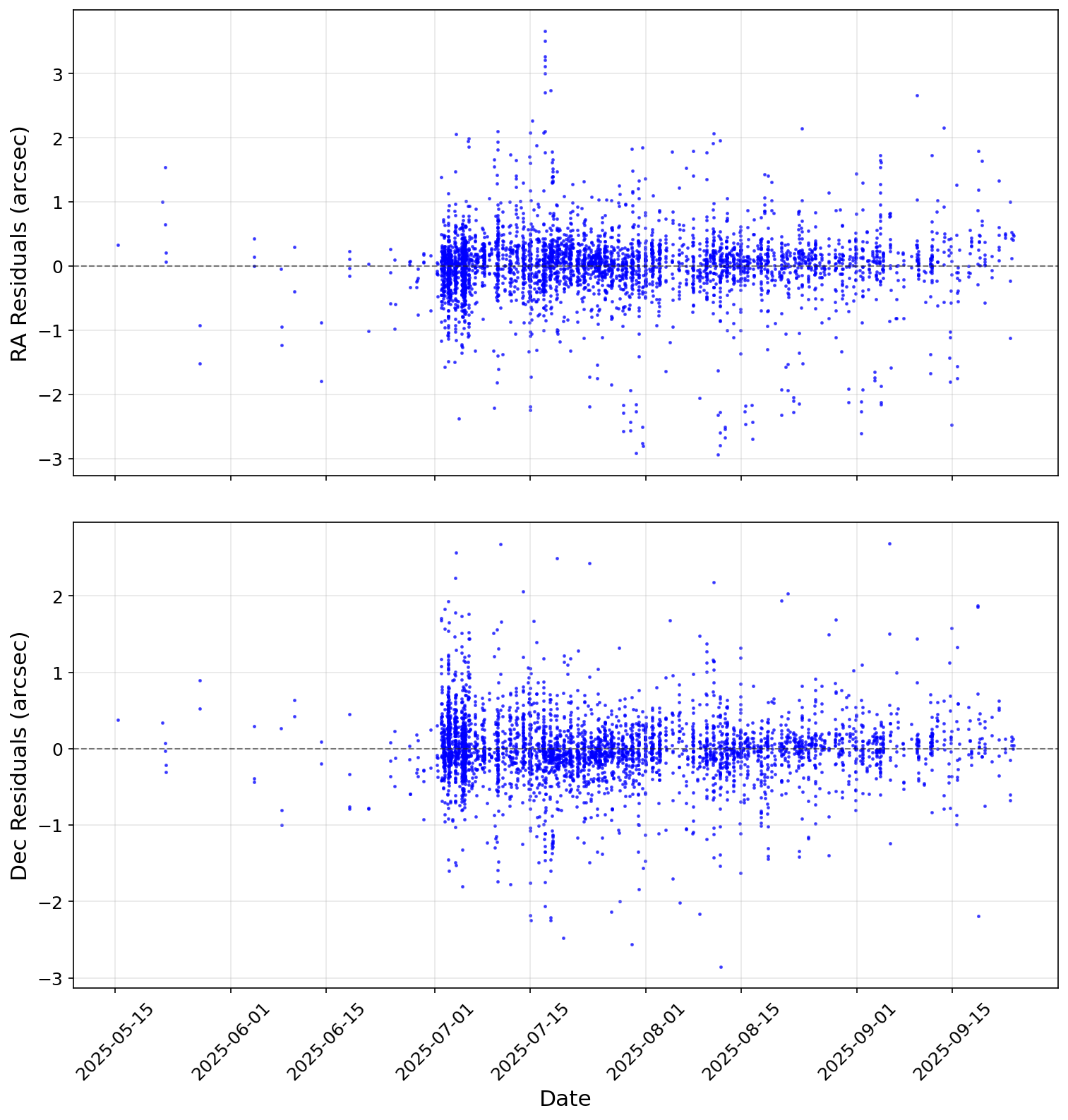}
    \caption{Astrometric residual analysis showing the difference between observed and predicted sky positions between mid-May and mid-September 2025. Top and bottom panels show RA and Dec residuals respectively. \textbf{Left panel:} Binned residuals averaged over 8 equal-duration time bins with error bars showing standard error of the mean. \textbf{Right panel:} Individual residual measurements for all 4,022 observations.}
    \label{fig:residuals_combined}
\end{figure*}

Figure~\ref{fig:residuals_combined} shows the RA and Dec residuals over the 4.5 month period. 

\section{Discussion}

The theoretical model assumes a trajectory dictated just by gravity. Any unmodeled non-gravitational forces (outgassing, radiation pressure, Yarkovsky effect) would appear as systematic residuals that could evolve with heliocentric distance and solar illumination conditions.


The RA and Dec residuals are consistent with zero between July 1 and September 23, 2025 to within $0.028^{\prime\prime}$. The residuals are unlikely to be affected by a systematic shift in the centroid of light owing to the development of a cometary bright spot away from the nucleus of 3I/ATLAS. The highest-resolution image of 3I/ATLAS, taken by HST on July 21, 2025, showed an optically-thin anti-tail in the direction of the Sun~\citep{Jewitt,Keto} which persisted during July and most of August. On August 27, 2025, imaging by the Gemini South telescope revealed the growth of a faint tail from 3I/ATLAS away from the Sun\footnote{\url{ https://noirlab.edu/public/news/noirlab2525/}}. Throughout the entire period, the brightest point remained centered on the nucleus.

Over a short time interval $\Delta t$, a non-gravitational acceleration $a_{\rm NG}$ results in a positional residual that grows as $\sim {1\over 2} a_{\rm NG}
 (\Delta t)^2$. Given that the distance of 3I/ATLAS from Earth in mid-August was $\sim 2.6~{\rm au}$~\citep{Lisse}, we use the measured RA residuals to derive an upper limit  of $a_{\rm NG}\lesssim 10^{-10}~{\rm au~d^{-2}}= 2\times 10^{-7} {\rm cm~s^{-2}}$. 

The momentum-balance equation implies,
\begin{equation}
M =-{{\dot M} v_{\rm ej} \over a_{\rm NG}} ,
\label{rocket}
\end{equation}
for an object of mass $M$ with an excess mass-loss rate from its Sun-facing side of ${\dot M}$ and an outflow ejection speed relative to the object’s surface of $v_{\rm ej}$. 

JWST data taken on August 6, 2025~\citep{Cordiner} imply a total ${\dot M}=-150~{\rm kg~s^{-1}}$ and $v_{\rm ej}\sim 0.44~{\rm km~s^{-1}}$. Substituting these values and the inferred $a_{\rm NG}$  into equation~(\ref{rocket}), yields an object mass of $M\gtrsim 3.3\times 10^{16}~{\rm g}$, corresponding to a nucleus diameter $D\gtrsim 5~{\rm km}$ for a bulk density of $\sim 0.5~{\rm g~cm^{-3}}$, similar to comets like 1P/Halley \citep{SosaFernandez2008}, 9P/Tempel 1 \citep{Veverka2006}, 81P/Wild 2 \citep{GutiérrezDavidsson2006} or 67P/Churyumov–Gerasimenko \citep{Preusker2017}. This lower limit on the diameter of 3I/ATLAS is close to the upper bound of the range of 0.44$-$5.6~km inferred from the HST data~\citep{Jewitt}, and is an order of magnitude larger than the estimated diameter of the interstellar comet 2I/Borisov~\citep{JewittSeligman}.  

A large nucleus mass for 3I/ATLAS exacerbates the tension between the interstellar mass reservoir of rocky material ejected from planetary systems and the implied detection rate of interstellar objects of its size~\citep{Loeb}. 

The passage of 3I/ATLAS near Mars on October 3, 2025 will allow the HiRISE camera onboard the Mars Reconnaissance Orbiter\footnote{\url{https://www.uahirise.org/}} to achieve a pixel resolution of $\sim 30~{\rm km}$. The amount of sunlight reflected from the brightest pixel in the HiRISE image will constrain further the nucleus surface area for an assumed albedo value. Additional constraints can be obtained on March 16, 2026, when 3I/ATLAS will pass in proximity to the Juno spacecraft near Jupiter~\citep{Loeb2}.

\bigskip
\noindent{\bf Acknowledgements}. 
This work was supported in part by the Galileo Project at Harvard University.

\bibliography{references}{}
\bibliographystyle{aasjournalv7}

\end{document}